\documentclass[conference]{IEEEtran}

\usepackage{bbold}
\usepackage{amsmath}
\usepackage{graphicx}
\usepackage{units}
\usepackage{bm}
\usepackage{multirow}
\usepackage{epstopdf}
\usepackage{url}

\usepackage{fancyhdr}
\begin{document}

\title{A Locational Price for Power Injection Fluctuations of Variable Generation and Load}

\author{\IEEEauthorblockN{Adria E. Brooks}
\IEEEauthorblockA{Department of Electrical and\\Computer Engineering\\
University of Wisconsin-Madison\\
Email: brooks7@wisc.edu}
\and
\IEEEauthorblockN{Bernard C. Lesieutre}
\IEEEauthorblockA{Department of Electrical and\\Computer Engineering\\
University of Wisconsin-Madison\\
Email: lesieutre@wisc.edu}
}

\maketitle

% have to add this command to get fancyfoot on first page
\thispagestyle{fancy}

\begin{abstract}
In this paper we calculate the incremental system production cost
associated with a measure of locational power injection uncertainty
that can be interpreted as a locational price for tracking power
fluctuations. This ``Locational Price of Variability (LPV)'' can be used to allocate charges for regulation
reserves by location, and hence, can also be used to value distributed 
energy storage employed to mitigate such fluctuations. We consider policy
changes that could enable the implementation of the LPV.

\end{abstract}
\section*{Nomenclature}
\begin{table}[h] \centering
\begin{tabular}{cl}
	\multicolumn{2}{l}{\textbf{Variables}}  \\
	$P_g$ & 	Scheduled conventional power generation \\
	$A$ &	AGC-based regulation reserve capacity \\
	$a$ &	Required AGC-based regulation reserves \\
	$\beta$ &	AGC-based regulation reserve participation factor \\
	\multicolumn{2}{l}{\textbf{Sets and Indices}}  \\
	$N$ &	Total number of all system buses \\
	$i, k, m$ &	Indices for buses \\
	$j$ & 	Index for transmission lines \\
	\multicolumn{2}{l}{\textbf{System Parameters}}  \\
	$C_g$ &	Conventional generator operating costs \\
	$C_A$ &	AGC-based regulation reserve capacity costs \\
	$P_g^{min}$ &	Minimum conventional generator limit \\
	$P_g^{max}$ &	Maximum conventional generator limit \\
	$P_{flow}$ &	Power flow on transmission lines \\
	$P_{br}^{max}$ &	Transmission line limit \\
	$P_{ND}$ &	Scheduled non-dispatchable power injections \\
	$\Delta P_{ND}$ &	Variability in scheduled non-dispatchable power injections \\
	$\Gamma$ &	Generation shift factors \\
	$d$ &	Generation weighting matrix \\
	$\epsilon_{a}$ &	Allowable violation probability for regulation reserves \\
	$\epsilon_{br}$ &	Allowable violation probability for transmission lines \\
	$\sigma$ &	Standard deviation of distribution \\
	\multicolumn{2}{l}{\textbf{Acronyms and Abbreviations}}  \\
	ACE &	Area control error \\
	AGC &	Automatic generation control \\
	LMP &	Locational marginal price \\
	LPV &	Locational price of variability \\
	FTR &	Financial transmission rights
\end{tabular}
\end{table}
\section{Introduction}
The widespread and distributed use of renewable energy sources such as wind and solar power has naturally led to increased variation in power injections as
the fluctuations in these sources add to fluctuation in load. This is manifest on many time-scales, from short-term load balancing (seconds-minutes), medium
term load tracking through daily ramps (minutes-hour), to long term capacity considerations (hours). See \cite{denholm:2011} for an illustrative plot of
increased variability due to wind generation in Texas, and an analysis of the generation flexibility and energy storage needs to support a very high profile
of wind generation. In addition to the discussion of the problems with increased ramp rates due to wind generation in Texas, there is considerable concern
about the ramp rates associated with solar power in California, i.e. the famous ``Duck Curve'' \cite{CAISO:duckcurve}. The dramatic change in solar output (sunrise/sunset)
necessitates very sharp changes in conventional generation in the absence of large amounts of energy storage. Similarly, but on a smaller scale, short-term
power fluctuations due to loads are exacerbated by fluctuation in distributed renewable generation (wind/solar). This paper focuses on these faster
time-scale variations associated with short-term balancing. We propose to calculate a locational price associated with short-term power variations that could be
applied uniformly to variable loads and generation, and that can serve as an incentive to support energy storage technologies to smooth power
generation/use.

In order to maintain system reliability, conventional generators under Automatic Generation Control (AGC) are typically employed for short-term load
tracking using an Area Control Error (ACE) signal. In systems operating with electricity markets, this power-balancing regulation action is treated as an
ancillary service, the costs of which are shared among participants. We suggest here that the costs of regulation services could be allocated in a manner
consistent with how the load/generation fluctuations affect total system marginal cost. In particular, load tracking controllable generation must hold some capacity
in reserve for tracking purposes, effectively reducing their nominal dispatch range. Also line-flow limits can be reduced to accommodate uncertainty in
resulting flows. We emphasize that the impact on system cost is more than the incremental cost of power balancing - which should average to near zero for
regulation. Rather, accommodating the variations will impact the nominal system dispatch and operation point, at a cost.

We calculate a price as the incremental cost associated with a measure of variability at each location. This requires the use of a probabilistic optimal
power flow to capture the effect of variable loads/generation. For the purposes of this paper, we adopt the modeling approach used in \cite{roald:2013}
in which generator limits and line flow limits are cast as chance constraints. Similar to \cite{ding:2016} and \cite{roald:2017} we include AGC-based regulation reserves in the model. In this framework the probability of exceeding a limit is imposed to be less than some user-specified value. This allows the use of a Gaussian distribution for the uncertainty in power, and offers the standard deviation as a convenient measure. We extend the analysis in that paper by considering load variations in addition to renewable generation. Then at each bus we can calculate a sensitivity of system cost to uncertainty in power injection. Using this as a price provides an incentive to reduce short-term power fluctuations. If energy storage is considered for this purpose, the price helps provide a specific value
for energy storage.

Unlike previous work, we consider variability of loads in addition to variable-resource renewable energy. We allow the optimization model to determine the amount of AGC-based regulation capacity to reserve instead of pre-assigning it. Other differences in our model include the use of a DC power flow instead of AC, we do not distinguish between up and down regulation reserves, nor do we consider other reserves or security constraints.

We mention that we introduced this conceptual approach in \cite{PECI:2015}, and that this paper corrects and improves upon that paper. Specifically in
\cite{PECI:2015} we used an overly-complex sensitivity model the results of which, we determined later, did not match empirically-derived sensitivities. In
this paper we provide a more direct sensitivity calculation that is consistent with empirically calculated sensitivities.

The modeling and results applied to a 14-bus system are presented in the following sections, and we conclude with a discussion of practical implementation.

\section{Methodology}
We define both the standard DC optimal power flow (DCOPF) and a DCOPF problem which considers operating reserves for conventional, dispatchable generators. We then extend the DCOPF with operating reserves to include probabilistic constraints on the AGC-based regulation reserves and transmission line limits to accommodate power injection variability. Though the energy generation market would not be used for short-term balancing in grid systems with ancillary service markets, we present the chance constrained DCOPF without regulation for a comparison.

%%%% SUB SECTION %%%%
\subsection{Standard DC Optimal Power Flow}

%% SUB SUB SECTION %%
\subsubsection{DCOPF with operating reserves}

In the standard DCOPF problem, conventional power generators \bm{$P_{g}$}
make up all scheduled non-dispatchable power loads and generation
\bm{$P_{ND}$}. Non-dispatchable generation include privately owned
distributed energy resources and variable-resource renewable generation
systems that the grid system operator cannot control. Operating reserves
are used to track the variability in the non-dispatchable power injections
\bm{$\Delta P_{ND}$}. This variability can be the result of fluctuating
demand or variable-resource generation. In order to accommodate this
variability, system operators request that several conventional generators
reserve some of their generation capacity for AGC regulation reserves
\bm{$A$}. The amount of actual generation used to track power
fluctuations \bm{$a$} is less than the capacity \bm{$A$} reserved for this purpose.

Equations (\ref{eq:genReg_objective}) - (\ref{eq:genReg_Pflow_exp})
  describe the DCOPF with operating reserves framed as a linear program.
  The objective (\ref{eq:genReg_objective}) is to minimize the costs of
  power generation \bm{$C_g^T P_g$} and reserve reserve capacity \bm{$C_A^T A$}. Reserve generation provided by a single generator \bm{$a_i$} makes up some portion of the total variability as described by the participation factor \bm{$\beta_i$}, defined in equation (\ref{eq:genReg_a}).

The generator and transmission line limits are described by
\bm{$P_{g}^{min}$}, \bm{$P_{g}^{max}$} and \bm{$P_{br}^{max}$}. Power flow
on the system transmission lines is defined by all power injections in the system,
   including both regulation reserves and variability, as shown in
   (\ref{eq:genReg_Pflow}). Here \bm{$\Gamma_{ji}$} are the generation
   shift factors which describe the change in power flow based on a change
   in power injection. We note that the form of shift factors generally
   depends on a choice of slack bus, or distributed slack, while the results of
   this optimization do not.  Any consistent shift factor representation
   will yield the same result. The base power generation profile,
   regulation reserve capacities, and the allocation of AGC weights are
   determined by the optimization problem. 

\begin{eqnarray}
	& \displaystyle \min_{P_{g},A} \; C_g^T P_g + C_A^T A  \label{eq:genReg_objective} \\
	\nonumber \\		% provide some space 
	\textrm{s.t. } & \sum_i^N P_{g,i} = - \sum_i^N P_{ND,i}	\\
	& P_{g}^{min} \le P_{g} - A	\label{eq:genReg_genMin}   \\
	& P_{g} + A \le P_{g}^{max}	\label{eq:genReg_genMax}   \\
	& 0 \le \beta  \\
	& \sum_i^N \beta_i = 1     \\
	& a \le A     \label{eq:genReg_a_con}  \\
	& |P_{flow}| \le P_{br}^{max}   \label{eq:genReg_Pflow_con}  \\
	\nonumber \\      % provide some space 
	\textrm{where} & a_i = - \beta_i \; \left( \sum_i^N \Delta P_{ND,i} \right)    \label{eq:genReg_a} \\
	& P_{flow,j} = \Gamma_j \; \left[P_g + a + P_{ND} + \Delta P_{ND} \right]   	\label{eq:genReg_Pflow}
\end{eqnarray}

Equation (\ref{eq:genReg_Pflow_exp}) expands
   (\ref{eq:genReg_Pflow}) by substituting for (\ref{eq:genReg_a}).
\begin{flalign}
	P_{flow,j} = & \sum_i^N \Gamma_{ji} \left( P_{g,i} + P_{ND,i} \right) +   \nonumber \\
	& \sum_i^N \Gamma_{ji} \left( \left( 1 - \beta_i \right) \Delta P_{ND,i} - \beta_i \sum_{k\neq i}^N \Delta P_{ND,k} \right)  \label{eq:genReg_Pflow_exp}
\end{flalign}

%% SUB SUB SECTION %%
\subsubsection{DCOPF without operating reserves}

In this simplified DCOPF problem without operating reserves, all variability in non-dispatchable power is made up by the conventional generators. This problem is defined in (\ref{eq:gen_objective}) -- (\ref{eq:gen_Pflow}) as a linear program. The objective function and decision variables have been reduced to only consider the conventional power generation and the associated production costs. Power flow on the transmission lines has been similarly simplified to only include generated power.

The generation weighting matrix \bm{$d$} in equation (\ref{eq:d}) describes how power injection variability at any bus will be distributed among all conventional generators in the system. Generation shift factors are calculated using the method described in \cite{zimmerman:2011} for the \bm{$d$} elements corresponding to a single load bus, which define the distributed slack bus weighting elements. Though the slack bus configuration (distributed versus reference) does not affect the dispatch solution provided by the DCOPF with operating reserves problem, it does affect the dispatch in this problem formulation.

\begin{eqnarray}
	& \displaystyle \min_{P_{g}} \; C_{g}^T P_{g} \label{eq:gen_objective} \\
	\nonumber \\		% provide some space 
	\textrm{s.t. } & \sum_i^N P_{g,i} = -\sum_i^N P_{ND,i}	  \\
	& P_{g}^{min} \le P_{g}	 \label{eq:gen_Pmin_con} \\
	& P_{g} \le P_{g}^{max}	 \label{eq:gen_Pmax_con} \\
	& |P_{flow}| \le P_{br}^{max} \label{eq:gen_Pflow_con} \\
	\nonumber \\      %provide some space
	\textrm{where} & P_{flow} = \Gamma_{ji} P_g	\label{eq:gen_Pflow}
\end{eqnarray}
\begin{equation}
	d_{mi} = \begin{cases}
		\frac{P_{g,m}^{max}}{\sum_{k\neq i}^{N} P_{g,k}^{max}}, & m \neq i	 \\[0.1in]
		0, & m=1
		\end{cases}	\label{eq:d}
\end{equation}

%%%% SUB SECTION %%%%
\subsection{Probabilistic Constraints}

%% SUB SUB SECTION %%
\subsubsection{DCOPF with operating reserves}
We extend the chance constraint approach developed by Roald, et al. \cite{roald:2013} to accommodate power injection variability at any bus given optimal power flow with regulation reserves. Power injection variations further constrain both the power generation and transmission limits on the system. The probability that the required regulation reserve \bm{$a$} is within the scheduled reservation capacity \bm{$A$} is less than one by some acceptable violation probability \bm{$\epsilon_a$}. Similarly, the probability that the power flow on transmission lines including variability is within the branch limits is less than one by \bm{$\epsilon_{br}$}. These probabilities are described in (\ref{eq:genReg_ProbA}) -- (\ref{eq:genReg_ProbBrMax}).
\begin{eqnarray}
	& \mathbb{P} \{ a \le A \} \ge 1-\epsilon_a     \label{eq:genReg_ProbA} \\
	& \mathbb{P} \{ |P_{flow}| \le P_{br}^{max} \} \ge 1-\epsilon_{br}	\label{eq:genReg_ProbBrMax}
\end{eqnarray}

To convert these probabilistic forms of the constraints into deterministic forms compatible with the DCOPF problem, we model (\ref{eq:genReg_ProbA}) and (\ref{eq:genReg_ProbBrMax}) as Gaussian normal distributions. In this way, a measure of the fluctuation from scheduled non-dispatchable power \bm{$\Delta P_{ND}$} is taken as the standard deviation of the distribution \bm{$\sigma$}. The resulting cumulative distribution function is defined in equation (\ref{eq:Gauss}). We solve (\ref{eq:Gauss}) for a given generic violation probability \bm{$\epsilon$} to obtain \bm{$x_{\epsilon}$}, as shown in (\ref{eq:Xepsilon}). The analytical reformulation of constraints (\ref{eq:genReg_ProbA}) and (\ref{eq:genReg_ProbBrMax}) are given in (\ref{eq:genReg_a_con2}) and (\ref{eq:genReg_Pflow_con2}), after substituting for \bm{$a_i$} using equation (\ref{eq:genReg_a}). These reformulations are used to replace the standard DCOPF constraints (\ref{eq:genReg_a_con}) and (\ref{eq:genReg_Pflow_con}).

Note we do not include a probabilistic constraint directly on the generator limits as defined in (\ref{eq:genReg_genMin}) and (\ref{eq:genReg_genMax}). We believe these limits will be treated as hard limits in practice, at least as far as AGC regulation reserve is concerned. Constraint (\ref{eq:genReg_ProbA}) represents instead the probability that enough AGC-based regulation capacity will be reserved to meet the cumulative variability.
\begin{eqnarray}
	\Phi\left(x_\epsilon\right) & = 1 - \epsilon = \int_{-\infty}^{x_{\epsilon}} \frac{1}{\sqrt{2\pi}} \; e^{\frac{-x^2}{2}} \; dx \label{eq:Gauss} \\
	x_{\epsilon} & = \sqrt{2} \; \left[ \textrm{erf}^{-1} \left(2 \left( 1-\epsilon \right) -1 \right) \right]  \label{eq:Xepsilon}
\end{eqnarray}
\begin{equation}
	-x_{\epsilon,a} \beta_i \sqrt{ \sum_i^N \sigma_i} \le A_i   \label{eq:genReg_a_con2} 
\end{equation}
\begin{flalign}
	& \Bigg| \sum_i^N \Gamma_{ji} P_{g,i} + \sum_i^N \Gamma_{ji} P_{ND,i} +   \nonumber \\
	& x_{\epsilon,br} \sqrt{ \sum_i^N \Gamma_{ji}^2 \left( 1-\beta_i \right)^2 \sigma_i^2 - \sum_i^N \Gamma_{ji}^2 \beta_i^2 \left(\sum_{k \neq i}^N \sigma_k^2 \right)} \, \Bigg| \le P_{br,j}^{max}       \label{eq:genReg_Pflow_con2} 
\end{flalign}

%% SUB SUB SECTION %%
\subsubsection{DCOPF without operating reserves}
In the DCOPF problem without regulation reserves, constraints (\ref{eq:gen_Pmin_con}) -- (\ref{eq:gen_Pflow_con}) are similarly modified to probabilistic forms to accommodate power injection variability in (\ref{eq:ProbGenMin}) -- (\ref{eq:ProbBrMax}). Using the same Gaussian distribution approach, these are converted to their deterministic forms (\ref{eq:gen_Pmin_con2}) -- (\ref{eq:gen_Pflow_con2}), where the deviation at every bus \bm{$\sigma_i$} is weighted by either \bm{$d_{mi}$} or the generation shift factor \bm{$\Gamma_{ji}$}.

\begin{eqnarray}
	& \mathbb{P} \left[ P_{g} + \Delta P_{ND} \ge P_{g}^{min} \right] \ge 1-\epsilon \label{eq:ProbGenMin} \\
	& \mathbb{P} \left[ P_{g} + \Delta P_{ND} \le P_{g}^{max} \right] \ge 1-\epsilon \label{eq:ProbGenMax} \\
	& \mathbb{P} \left[ | \Gamma_{ji} (P_{g} + \Delta P_{ND}) | \le P_{br}^{max} \right] \ge 1-\epsilon	\label{eq:ProbBrMax}
\end{eqnarray}
\begin{eqnarray}
	& P_{gen} \ge P_{g}^{min} + x_{\epsilon} \sqrt{\sum_{i}^{N} d_{mi}^{2} \; \sigma_{i}^{2}}   \label{eq:gen_Pmin_con2}  \\
	& P_{gen} \le P_{g}^{max} - x_{\epsilon} \sqrt{\sum_{i}^{N} d_{mi}^{2} \; \sigma_{i}^{2}}    \label{eq:gen_Pmax_con2}  \\
	& |\Gamma_{ji} P_{g}| \le P_{br}^{max} - x_{\epsilon} \sqrt{\sum_{i}^{N} \Gamma_{ji}^{2} \; \sigma_{i}^{2}}       \label{eq:gen_Pflow_con2}
\end{eqnarray}

%%%% SUBSECTION %%%%
\subsection{Locational Price of Variability}

The price of variability is defined as the change in total system cost due
to a change in power injection variability at a given bus. This can be
found either empirically by perturbing the power injection variability at
each bus individually or analytically by calculating the partial
derivative of the Lagrangian function with respect to the standard
deviation \bm{$\sigma$}. The Lagrangian for the reformulated chance
constrained problem is given in equation (\ref{eq:lagrangian}). The
locational marginal price is the sensitivity of the Lagrangian with
respect to the scheduled non-dispatchable load at every bus. This is shown
in equation (\ref{eq:LMP}). We define the locational price of variability
(LPV) equivalently as the sensitivity of the Lagrangian with respect to
standard deviation, as shown in (\ref{eq:LPV}).

As presented here, the calculation results in the dispatch of generation and regulation reserves for conventional generators, and locational prices for both energy and variability. Because this calculation occurs before real-time, it relies on an estimate of variability (mathematical uncertainty), not a measured variability.

\setlength{\arraycolsep}{0.0em}
\begin{eqnarray}
	\mathcal{L}&{}={}& \left( C_{g}^T P_{g} + C_{a}^T A \right) + \lambda_1^T \left( \sum_i^N P_{g,i} + \sum_i^N P_{ND,i} \right)		\nonumber \\
	& + \lambda_2^T & \left( P_{g}^{min} - P_{g} + A \right) + \lambda_3^T \left( P_{g} + A - P_{g}^{max} \right)	\nonumber \\
	& + \lambda_4^T & \left( - \beta \right) + \lambda_5^T \left( \sum_i^N \beta_i - 1 \right)	\nonumber \\
	& + \lambda_6^T & \Bigg( -x_{\epsilon,a} \, \beta_i \sqrt{ \sum_i^N \sigma_i^2 } -A_i \Bigg)	\nonumber \\
	& + \lambda_7^T & \Bigg( -P_{br,j}^{max} + \sum_i^N \Gamma_{ji} P_{g,i} + \sum_i^N \Gamma_{ji} P_{ND,i} +	\nonumber \\
	& & x_{\epsilon,br} \sqrt{ \sum_i^N \Gamma_{ji}^2 \left( 1 - \beta_i \right)^2 \sigma_i^2 - \sum_i^N \Gamma_{ji}^2 \beta_i^2 \bigg(\sum_{k \neq i}^N \sigma_k^2 \bigg)} \, \Bigg)  \nonumber \\
	& + \lambda_8^T & \Bigg( -P_{br,j}^{max} - \sum_i^N \Gamma_{ji} P_{g,i} - \sum_i^N \Gamma_{ji} P_{ND,i} - 	\nonumber \\
	& & x_{\epsilon,br} \sqrt{ \sum_i^N \Gamma_{ji}^2 \left( 1 - \beta_i \right)^2 \sigma_i^2 - \sum_i^N \Gamma_{ji}^2 \beta_i^2 \bigg(\sum_{k \neq i}^N \sigma_k^2 \bigg)} \, \Bigg) \nonumber \\
        \label{eq:lagrangian}
\end{eqnarray}
\setlength{\arraycolsep}{5pt}

\setlength{\arraycolsep}{0.0em}
\begin{flalign}
	& LMP = -\frac{\partial \mathcal{L}}{\partial P_{ND}} = \; - \lambda_1^T - \lambda_7^T \bigg( \sum_i^N \Gamma_{ji} \bigg) +  \lambda_8^T \bigg( \sum_i^N \Gamma_{ji} \bigg)  	 \label{eq:LMP} \\
	& LPV = \frac{\partial \mathcal{L}}{\partial \sigma} = \lambda_6^T \bigg( -x_{\epsilon,a} \, \beta_i \frac{\sigma_i}{\sqrt{\sum_i^N \sigma_i^2}} \bigg)   \nonumber \\
	& + \lambda_7^T \Bigg( x_{\epsilon,br} \frac{ \sum_i^N \sigma_i \Big( \Gamma_{ji}^2 (1-\beta_i)^2 - \sum_{k \neq i}^N \Gamma_{jk}^2 \beta_k^2 \Big) }{ \sqrt{\sum_i^N \Gamma_{ji}^2 (1-\beta_i)^2 \sigma_i^2 - \sum_i^N \Gamma_{ji}^2 \beta_i^2 ( \sum_{k \neq i}^N \sigma_k^2 )} } \Bigg)   \nonumber \\
	& - \lambda_8^T \Bigg( x_{\epsilon,br} \frac{ \sum_i^N \sigma_i \Big( \Gamma_{ji}^2 (1-\beta_i)^2 - \sum_{k \neq i}^N \Gamma_{jk}^2 \beta_k^2 \Big) }{ \sqrt{\sum_i^N \Gamma_{ji}^2 (1-\beta_i)^2 \sigma_i^2 - \sum_i^N \Gamma_{ji}^2 \beta_i^2 ( \sum_{k \neq i}^N \sigma_k^2 )} } \Bigg)						    \label{eq:LPV}  
\end{flalign}
\setlength{\arraycolsep}{5pt}

We end this section by observing that the optimization problem with the probabilistic constraints is not a linear program, even though we began with a linear DCOPF description. It is not the purpose of this paper to investigate the range of nonlinear optimization tools available, and we use Matlab-based tools here. The results are presented in the next section.

\section{Results}

%%%% SUB SECTION %%%%
\subsection{Case System Characteristics}
\begin{figure}[h]	
\centering
\includegraphics[width=3.7in]{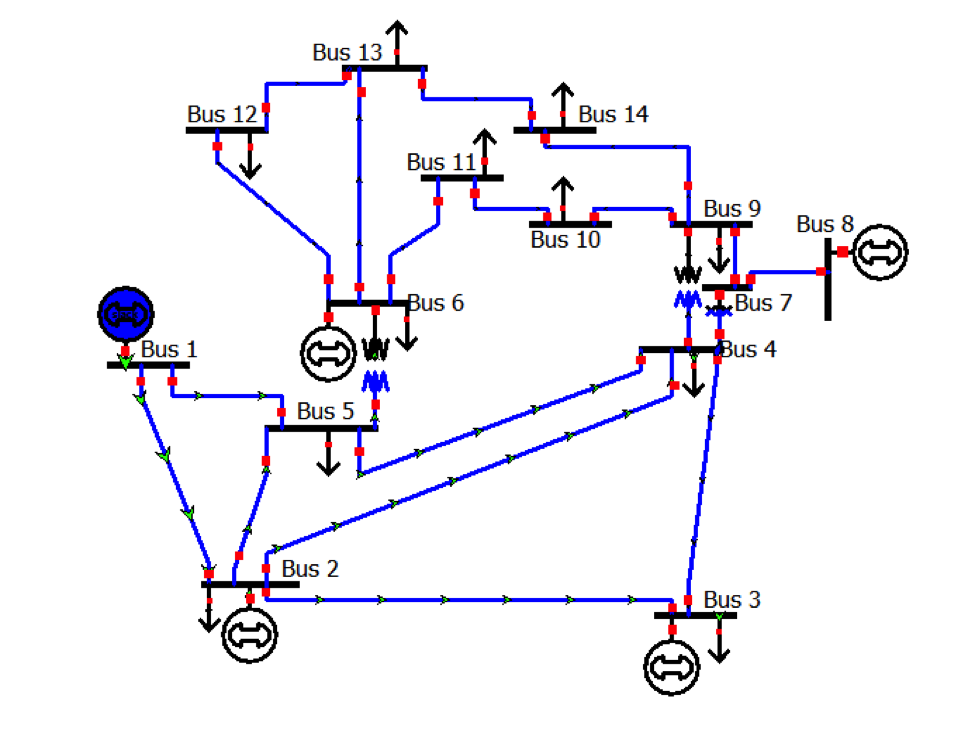}
\caption{IEEE 14 bus test system diagram \cite{ICSEG:14bus}.}  	\label{fig:14bus}
\end{figure}

The system studied here is a modified IEEE 14-bus system, shown in Fig.
\ref{fig:14bus}. There are fourteen buses, twenty branches, five
conventional generators and one wind generator at bus 14, modeled here as
a negative load. We assume the standard deviation associated with power
injection variability is 2\% for all loads and 10\% for the wind
generator. The system generation and load information are given in Table
\ref{table:bus}. The transmission branch data is unchanged from the IEEE
14-bus system, except the transmission
line between buses 2 and 5 is forced to be binding at a limit of 100
MVA. We assume an acceptable violation probability of 1\% for both the
regulation \bm{$\epsilon_a$} and line limits \bm{$\epsilon_{br}$}.

\begin{table}[h] \centering
\caption{System Bus Information} \label{table:bus}
\begin{tabular}{c|cccccc}
	Bus & $P_{load}$ & $\sigma$ & $P_{gen}^{min}$ & $P_{gen}^{max}$ & $C_{gen}$  & $C_{reg}$ \\
	No. & (MW) & (MW) & (MW) & (MW) & (\$/MWh) & (\$/MWh) \\  \hline
	1 & 34 & 0.68 & 15 & 332.4 & 21 & 16\\
	2 & 12 & 0.24 & 15 & 140.0 & 20 & 11\\
	3 & 9 & 0.18 & 15 & 100.0 & 35 & 13\\
	4 & 85 & 1.70 & & &  & \\	
	5 & 60 & 1.20 & & &  & \\
	6 & 22 & 0.44 & 15 & 100.0 & 39 & 12\\
	7 & 103 & 2.06 & & & & \\
	8 & 30 & 0.60 & 15 & 100.0 & 40 & 14\\
	9 & 61 & 1.22 & & & & \\
	10 & 74 & 1.48 & & & & \\
	11 & 15 & 0.30 & & & & \\
	12 & 57 & 1.14 & & & & \\
	13 & 66 & 1.32 & & & & \\
	14 & -50 & 5.00 & & & &
\end{tabular}
\end{table}

%%%% SUB SECTION %%%%
\subsection{Results}

The results of the chance constrained  DCOPF both with and without
regulation reserves are compared in Table \ref{table:econDisp}. The
economic generator dispatches and overall system costs solved for this
case are given. Table \ref{table:LMP} shows the calculated locational
marginal prices and locational prices of variability for the cases with
and without regulation reserves. The effects of the line constraint can be
observed in the generator dispatches. The generators at buses 1 and 6 are
dispatched at their maximum levels and the generators at buses 3 and 8 are
dispatched at their minimum, accounting for the AGC reserves. These
minimum dispatch generators result in lower LMPs than their offer prices.
We did not consider a unit committment with this model and it may be the
case that one or both of these generators may not be needed for the
least-cost disptach. If they are necessary for either energy or
regulation, then the difference would typically be made up via an out-of-market make whole
payment to these suppliers. Note that the generator
at bus 6 is more expensive than the generator at bus 3 with respect to its
energy offer price. The line constraint necessitates the
heavy use of this more expensive generator to supply the load in a portion
of the network.

The system cost of regulation reserves for this case is small in
comparison to the cost of power generation. The locational price of
variability is generally smaller than the locational marginal
prices for many buses where power injection variability is low. For a bus
with high variability, such as the wind generator at bus 14, power
injection variability costs nearly three quarters the price of
electricity. The larger load at bus 7 and its higher variability
contribute to its slightly higher LPV. Note that the more variable loads
and generators contribute more to the combined variance in overall power
fluctuations, leading to their greater impact on sensitivity studies. This
expectantly leads to higher prices at those locations.  The line constraint
also affects the LPVs.
Despite having low load variability, bus 1 also has a high
LPV. At this bus regulation reserves are comparatively expensive and the
generator is already operating at its generation capacity. The line
constraint impedes the paths to supply balancing power for fluctuations.
The prices at buses 2 and 5 are slightly elevated for similar reasons;
these buses are the terminals of the congested line.

%%%% SUB SECTION %%%%
\subsection{Significance}
We have proposed the calculation of locational prices of variability in
order to allocate the costs of AGC regulation to the locations whose
variabilities most impact the system.  Not surprisingly, the bus locations with
large variable power injections and locations which impact the constrained line
have larger prices. These prices provide an incentive to reduce
variations in power injections.  Consider the wind generator at bus 14.
As a generator it would receive revenue of 2039 (\$/hr) for the supply of
50 MW, and be charged 137 (\$/hr) for its variability, assuming its real-time
fluctuations were similar to the expected variations. This charge amounts
to almost 7 percent of its revenue. The assessment of such a charge would
almost certainly warrant a cost/benefit study for the purchase of energy storage
capability to reduce power fluctuations. The effect on the loads is similar
but smaller because we assumed smaller fluctuations. For example, the 66 MW
load at bus 13 would pay approximately 2769 (\$/hr) for energy and 10 (\$/hr)
for variability. The added charge is less than one half of one percent of
the energy charge.  Nevertheless, if some load had large
power fluctuations the LPV would also provide an incentive to consider energy
storage or other means to smooth the power usage profile.

\begin{table}[h] \centering
\caption{Economic Generator Dispatch} \label{table:econDisp}
\begin{tabular}{c|ccc|c}
	 & \multicolumn{3}{c|}{With CC Reserves} & Without CC Reserves \\ \hline
	Generator & $P_{g}$ & $A$ & $\beta$ & $P_{g}$ \\
	Bus No. & (MW) & (MW) & (pu) & (MW)   \\  \hline
	1 & 332.4 & 0 & 0 & 326.0\\
	2 & 108.5 & 0.08 & 0.006 & 105.1 \\
	3 & 15.0 & 0 & 0 & 17.0\\
	6 & 96.1 & 3.91 & 0.261 & 98.0 \\
	8 & 26.0 & 11.00 & 0.734 & 31.9 \\ \hline
	System Cost & {\multirow{2}{*}{14463}} & {\multirow{2}{*}{202}} & & {\multirow{2}{*}{14641}} \\
	(\$/hr) & & & & 
\end{tabular}
\end{table}
\begin{table}[h] \centering
\caption{Locational Marginal Pricing and Price of Variability} \label{table:LMP}
\begin{tabular}{c|cc|cc}
	 & \multicolumn{2}{c|}{With CC Reserves} & \multicolumn{2}{c}{Without CC Reserves} \\ \hline
	Bus & LMP & LPV & LMP & LPV \\
	No. & (\$/MWh) & (\$/MWh) & (\$/MWh) & (\$/MWh) 	 \\ \hline
	1 & 25.09 & 28.57 & 25.28 & 1.17 \\
	2 & 20.00 & 17.13 & 20.00 & 0.99 \\
	3 & 29.76 & 4.07 & 30.12 & 0.21 \\
	4 & 38.20 & 9.99 & 38.87 & 6.59 \\	
	5 & 44.27 & 10.11 & 45.16 & 10.41 \\
	6 & 42.29 & 2.75 & 43.11 & 3.06 \\
	7 & 39.29 & 11.20 & 40.00 & 9.40 \\
	8 & 39.29 & 3.26 & 40.00 & 2.95 \\
	9 & 39.87 & 6.54 & 40.61 & 6.06 \\
	10 & 40.30 & 7.96 & 41.05 & 7.81 \\
	11 & 41.28 & 1.69 & 42.06 & 1.81 \\
	12 & 42.10 & 6.97 & 42.91 & 7.63 \\
	13 & 41.95 & 7.94 & 42.76 & 8.66 \\
	14 & 40.78 & 27.35 & 41.55 & 28.15
\end{tabular}
\end{table}
\section{Conclusion}
In this paper we proposed and demonstrated a means to calculate a
locational price to apply to power injection fluctuations. Our model
expanded on our previous work by explicitly adding a model for AGC.  This
includes a market-based method for choosing generators to supply AGC in
the co-optimized energy and regulation cost function.  The probabilistic power
flow also determined the amount of regulation reserves that are required,
and the weights for the AGC controlled generators. The locational
price of variability was calculated as the sensitivity of system cost
to a measure of variability, in this case the standard deviation. We noted
that the locations with the highest LPVs were those with either highly
fluctuating power injections, or with power injections that impacted the
power flow on the constrained line.  We argued that the imposition of a
locational charge on variability would more fairly pay for resources
needed to balance load in real time. Instead of sharing the costs
uniformly, the non-dispatchable resources with fluctuations that most
affect system costs would be charged more for their variable power
injections.  These charges in turn provide an incentive to consider energy
storage technologies to smooth power fluctuations.  

Accepting the approach as reasonable, there are practical policy
questions to consider for implementation in a market. 
Importantly there is a potential gap between calculating the price based on uncertainties
(expectations of variability) and charging for the actual measured
fluctuations. The prices presented in Table III are based on
short-term expectations of nominal load, expectations of variations in
load and variable renewable generation.  The realizations of actual
 measured load,
generation, and fluctuations may differ from the expectations used to
calculate the prices. One could recalculate the prices ex-post,
however the cost of the decision on generator dispatch was already made.  There are
several possible resolutions to this issue.  If the history of
uncertainty projections proves to match observed fluctuations well, then
the method and charges can be applied as outlined in the paper. If however
there are significant differences then it would make sense to dispatch
generation and allocate reserves based on the best information available,
and then compute prices for variability ex-post. 

In either case --- using prices calculated with resource allocations or
prices calculated after variability is measured --- the amount charged to
non-dispatchable resources will likely differ from the costs of procuring
regulation reserves. This is the case even with perfect projections of
uncertain power injections. In our 14-bus example system, if we paid the
AGC generators the highest accepted price of 14 (\$/MWhr), then they
would receive in total 210 (\$/hr) for the allocated AGC capacity.  The
variable non-dispatchable loads and generation would pay in total 255
(\$/hr). In this case, the sources of variable power injections would by
paying 45 (\$/hr) more than the AGC resources would receive and some policy would be
required to handle the difference.  In energy markets it is typical for
loads to pay more than the the generators received due to congestion
costs. In our case example using LMPS to settle payments to generators
and from loads,
the difference in total payments will be 8580
(\$/hr). Congestion costs can be substantial and are settled using
Financial Transmission Rights (FTR), for which there is a separate market.
In the case of AGC, the amounts of power and the relative costs are so much
smaller than the energy market that it hardly warrants a complicated
structure to resolve the settlement.  We recommend the price profile be
used to determine the fair allocation of costs among participating entities
and pro-rate the charges to cover the costs of procuring balancing power
via regulation reserves. Alternatively, the small difference in payment for AGC regulation reserves could be included in the existing FTR market.

%Using the calculated price to allocate costs based on actual measured
%fluctuations, and for purposes of discussion, assuming the measured
%fluctuations are consistent with assumed uncertainties, we can
%compute the charges/payments for loads/generators. Focusing on the
%wind generator at bus 14, we calculate that it should receive a
%payment for generation equal to the product of the LMP and its nominal
%output of 50MW: 2,077.50 (\$/hr). Based on an observed sample standard
%deviation of 5WM and the price of variability, this unit should pay
%146.65 (\$/hr) for its contribution to regulation costs. By comparison,
%the cumultive load at bus 13 would expect to pay 1,634.16 (\$/hr) for
%delivered power plus 2.77 (\$/hr) due to regulation contribution. The
%application of the latter charge provides an incentive to at least
%determine the worth of employing energy storage to mitigate the
%variability costs. The price indirectly places a value on storage by
%location. A deeper analysis will be provided in the full paper. 

\section*{Acknowledgment}
%This material is based upon work supported by the National Science
%Foundation Graduate Research Fellowship Program under Grant No.
%DGE-1256259. Any opinions, findings, and conclusions or
%recommendations expressed in this material are those of the authors
%and do not necessarily reflect the views of the National Science
%Foundation. 

%This material is
%based upon work supported by the U.S. Department of Energy, Office of
%Science, Office of Advanced Scientific Computing Research, Applied
%Mathematics program under contract number DE-AC02-06CH11357 through
%the project "Multifaceted Mathematics Center for Complex Energy
%Systems" and under Argonne National Laboratory subcontract number
%3F-30222.

The authors gratefully acknowledge support from
the National Science
Foundation Graduate Research Fellowship Program under Grant No.
DGE-1256259, and
the U.S. Department of Energy, Office of
Science, Office of Advanced Scientific Computing Research, Applied
Mathematics program under contract number DE-AC02-06CH11357 through
the project "Multifaceted Mathematics Center for Complex Energy
Systems" and under Argonne National Laboratory subcontract number
3F-30222.

\bibliographystyle{IEEEtran}
\bibliography{ref}

\end{document}